# Comparison of Atmospheric Neutrino Flux Calculations at Low Energies

T.K. Gaisser[1], M. Honda[2], K. Kasahara[3], H. Lee[4],

S. Midorikawa[5], V. Naumov[6] and Todor Stanev[1]

[1]*Bartol Research Institute, University of Delaware, Newark, DE 19716, USA*

[2]*Inst. for Cosmic Ray Research, Univ. of Tokyo, Tanashi, Tokyo 188, Japan*

[3]*Faculty of Engineering, Kanagawa University, Yokohama 221, Japan*

[4]*Dept. of Physics, Chungnam National University, Daejeon 305-764, Korea*

[5]*Faculty of Engineering, Aomori University, Aomori 030, Japan*

[6]*INFN, Sezione di Firenze, Firenze I-50125, Italy*

We compare several different calculations of the atmospheric neutrino flux in the energy range relevant for contained neutrino interactions, and we identify the major sources of difference among the calculations. We find nothing that would affect the predicted ratio of $\nu_e/\nu_\mu$, which is nearly the same in all calculations. Significant differences in normalization arise primarily from different treatment of pion production by interactions of protons in the atmosphere. Different assumptions about the primary spectrum and treatment of the geomagnetic field are also of some importance.

Typeset Using *REVTEX*

# I. INTRODUCTION

Two deep underground detectors [1,2] observe a ratio of electrons to muons that is significantly different from what is expected from the spectra of $\nu_e$ and $\nu_\mu$ in the atmosphere. These two experiments use large volumes of water to detect Cherenkov radiation from charged particles that originate from interactions of neutrinos inside the detector. Together these two experiments have collected about 80% of the world's statistics of atmospheric neutrino interactions. Events with a single Cherenkov ring, which are mostly quasi-elastic, charged-current interactions of neutrinos, constitute the simplest and largest class of events in these detectors. The observed ratio of electron-like to muon-like events is significantly greater than expected from calculations.

A major component of the calculations (along with neutrino cross-sections and detector response) is the evaluation of fluxes of atmospheric neutrinos. Four sets of atmospheric neutrino spectra have been used in the past several years to interpret the measurements of interactions of $\nu_e$ ($\bar{\nu}_e$) and $\nu_\mu$ ($\bar{\nu}_\mu$) in underground detectors. All four flux calculations agree within a range of 5% for the flavor ratio of neutrinos with $0.4 \leq E_\nu \leq 1$ GeV [3], which is perhaps not surprising as most of the sources of uncertainty cancel in the calculation of this ratio. Much larger differences exist among the results for normalization and shape of the spectra, and these lead to ambiguities in the interpretation of the anomalous flavor ratio. For example, the calculation of Bugaev and Naumov (BN) [4] has a harder spectrum than the other calculations [5,6,7] (relatively less neutrinos below $E_\nu = 500$ MeV than above 1 GeV). Such a hard spectrum allows the suggestion [8] that the observed relative excess of electron-like events could be the result of the decay $p \rightarrow e^+ \nu \nu$. Even if spectra have the same shape, overall differences in normalization suggest different interpretations of the anomaly in terms of neutrino oscillations. A high normalization that agrees with the observed electron flux favors oscillations predominantly in the $\nu_\mu \leftrightarrow \nu_\tau$ sector [6,9] whereas a low normalization would suggest an oscillation that includes $\nu_e$. A calculation with a low normalization relative to which there is an excess of electrons could also be consistent with an interpretation in terms of a neutron-induced background masquerading as interactions of $\nu_e$ [10].

Of the four calculations we consider, three are completely independent of each other. The independent calculations are Honda et al. (HKHM) [6], Bugaev & Naumov (BN) [4] and Barr, Gaisser & Stanev (BGS) [5]. The work by Lee & Koh (LK) [7] uses a three-dimensional version of the model of hadronic interactions from BGS. LK also use the same primary spectrum as BGS. We have discovered several bugs in the implementation of the LK code. When these are removed, the results of LK are essentially the same as those of BGS in the absence of a geomagnetic cutoff. Moreover, since the transverse momentum of a neutrino from decay of a pion or muon is typically no more than 30 MeV, for the energies of interest here ($E_\nu > 200$ MeV) a three dimensional calculation is not necessary [11,12]. In what follows we therefore do not consider separately the calculation of LK.

The calculation of BGS is a Monte Carlo simulation made in two steps: first, cascades are generated for primary protons at a series of fixed primary energies over an appropriate range of zenith angles; second, the resulting yields of neutrinos, binned in $E_\nu$, are added together after weighting by the primary spectrum and geomagnetic cutoffs characteristic of each detector location, as calculated in Ref. [13], which neglected the effect of the penumbra of the Earth. This structure allows us to substitute other assumptions about primary spectra and composition and about geomagnetic cutoffs while keeping all other input the unchanged. Thus we can compare the sensitivity to these assumptions one-by-one in isolation. Hadron production in BGS is described in a model called TARGET [14] which is a parametrization of accelerator data for hadron nucleus collisions with emphasis on interaction energies around 20 GeV, which are most important for production of GeV neutrinos.

The calculation of BN is a semi-analytic integration of the atmospheric cascade equations in "straight-forward approximation" over the primary spectrum as modified by appropriate geomagnetic cutoff rigidities for protons and nuclei. Namely, BN used detailed tables for effective vertical cutoffs from Ref. [15] (corrected for the displacement of the geomagnetic poles with time) and a dipole-like relation for the directions different from vertical. In this approach, the penumbra structure, contribution of re-entrant albedo and direct influence of the geomagnetic field on the charged secondaries were neglected[1]. It has been estimated [19] that these effects are at the level of 10% or less for atmospheric neutrino fluxes averaged over reasonably wide solid angle bins.

The hadronic interaction model used by BN is an analytic parametrization of double-differential inclusive cross sections [18], which is based on great array of accelerator data and, according to [18], it applies at $p_N > 1$ GeV/$c$, $p_{\pi^\pm} > 150$ MeV/$c$ and $p_{K^\pm} > 300$ MeV/$c$. The comparison of the model with some data, which was not used when fitting its parameters, was presented in Ref. [16] (see also Ref. [17]). The exact inclusive kinematics was drawn on to make all necessary integrations. Other details of the BN calculation were described in Refs. [19,20,16].

The work of HKHM is a Monte Carlo calculation that includes a detailed treatment of the effect of the geomagnetic field. Cutoffs are calculated for each detector location by backtracing antiprotons through a map of the geomagnetic field. This procedure was also used by LK, and a similar analysis has recently been carried out by Lipari and Stanev [21]. This is the correct way to account for cutoffs because it includes the effects of trajectories that are forbidden because they intersect the surface of the Earth (penumbra). For the interactions above 5 GeV, HKHM use the subroutine packages FRITIOF version 1.6 [22] and JETSET version 6.3 [23]. At lower energy the algorithm NUCRIN [24] is used.

All calculations include the effect of muon polarization on the neutrinos from muon decay, following the remark of Volkova [25] who emphasized its importance in this context[2].

A quantitative comparison of the three independent calculations is made in Table 1. The first two blocks show the neutrino fluxes (normalized to BGS) in three ranges of energy, $0.4 < E_\nu < 1$, $1 < E_\nu < 2$, and $2 < E_\nu < 3$ GeV. The third block compares the neutrino ratios in the energy range $0.4 < E_\nu < 1$ GeV. We tabulate $R_{e/\mu} = (\nu_e + \frac{1}{3}\bar{\nu}_e)/(\nu_\mu + \frac{1}{3}\bar{\nu}_\mu)$ because the cross section for quasi-elastic interactions of antineutrinos is roughly one-third that of neutrinos in the low energy range relevant for single-ring contained events. Table 1 demonstrates the main differences between the different calculations. In addition to the overall normalization, there are also significant differences in the $\bar{\nu}_e/\nu_e$ ratio. The difference between BGS and HKHM, 0.89 vs. 0.84 averaged over the 0.4 to 1 GeV range, is actually quite big at neutrino energy below 500 MeV and disappears above 1 GeV.

---

[1]The last two effects were also neglected in the other calculations.

[2]The fluxes shown by BN in Ref. [4] do not include muon polarization, but they have since been corrected for this effect together with the effect of muon depolarization caused by muon energy loss [26,17].

We divide our discussion into three sections. We first compare the assumptions about the primary spectrum and about the geomagnetic cutoffs made in the three calculations. We then compare the treatment of hadronic interactions. We conclude with a brief discussion of the implications for interpretation of the measurements of contained neutrino interactions and the anomalous flavor ratio of neutrinos.

## II. PRIMARY SPECTRUM, COMPOSITION AND GEOMAGNETIC CUTOFFS

The cosmic ray beam consists of free protons and of nucleons bound in nuclei. The two must be distinguished because the primary spectrum and geomagnetic cutoffs depend on rigidity, while production of secondaries depends on energy per nucleon. At fixed rigidity, the momentum per nucleon for nuclei is half that for free protons to a good approximation.

BGS assume equal numbers of neutrons and protons for the bound nucleons and obtain the neutron yields from the proton yields by reversing all electric charges of mesons. This is exact for pions (by isospin symmetry) but introduces a small excess of charged kaons in interactions of neutrons. However, kaons contribute less than 10% of the neutrinos in the energy range relevant for contained events (E. Frank, private communication, 1993). A separate calculation for incident protons and neutrons has since been made [12] that has verified that this approximation has a negligible effect for the energy range relevant for contained interactions.

All bound nucleons are assumed to interact with the same starting point distribution as free nucleons; i.e., their cascades are simulated exactly as if they were free. This assumption can be derived for neutrinos (where correlations between different nucleons in the nucleus are irrelevant) within the framework of the Glauber multiple scattering picture of nucleus-nucleus collisions [27]. In contrast, BN treat nuclear interactions separately using the nucleus-nucleus cross sections to determine where the interactions occur [19,20].

Given the two-part structure of the BGS calculation, it is possible to check the effect of the different treatment of primary spectrum (including the model of nuclear interactions) separately from all other effects. We have done this by using the BN and the HKHM primary spectra and the BGS yields and geomagnetic cutoffs. Table 2 shows the effective differential spectrum of primary nucleons for BGS and BN at solar minimum. Replacing the BGS primary spectrum with that of BN decreases the neutrino flux by only about 5% below 1 GeV, leaves it nearly unchanged between 1 and 2 GeV and increases the calculated flux by about 5% for $2 < E_\nu < 3$ GeV. The neutrino flavor ratio also remains largely unchanged. Thus the differences in the primary spectrum and the treatment of nuclear projectiles are not the origin of the significant differences between these two calculations.

The primary spectra of BGS and HKHM averaged over the solar cycle are compared in Table 3. Use of the HKHM primary spectrum instead of that of BGS increases the neutrino flux in the energy range 0.4 – 1 GeV by about 7% and by about 12% in the 1 – 2 GeV range. The flux ratios again remain unchanged. In addition the BGS primary spectrum is significantly heavier, i.e. contains more bound nucleons, than either BN or HKHM spectra.

To check for the influence of the geomagnetic cutoff models, we replace the offset dipole model (with no penumbra) [13] used by BGS with the detailed treatment of the cutoffs used by HKHM. The flux of neutrinos between 0.4 and 1 GeV decreases by 10%. The geomagnetic effect is less important at higher energy. Note that the direction of this change is opposite to that caused by the different primary spectra used by BGS and HKHM, which was discussed above. We discuss this point further in the concluding section.

## III. TREATMENT OF HADRONIC INTERACTIONS

A useful way to compare the interaction models is to evaluate the spectrum-weighted moments (Z-factors) of the inclusive cross sections in the energy range relevant for the contained neutrino events. The Z-factors were not actually used in any of the calculations but they characterize the efficiency of the particle physics model to produce secondary particles in atmospheric cascades. The most important range of interaction energies for production of neutrinos with energies from 300 MeV to 3 GeV is $\sim 5 \leq E_N \leq 50$ GeV [30]. In this energy range cross sections do not scale so the Z-factors are energy-dependent:

$$Z_{p \to \pi^\pm} = \int_0^1 \mathrm{d}x \, x^{1.7} \frac{\mathrm{d}n_{\pi^\pm}(x, E_N)}{\mathrm{d}x},$$

where $x = E_\pi/E_N$, $E_N$ is the total energy of the incident nucleon in the lab system and $E_\pi$ is the energy of the produced pion.

The Z-factors are shown in Figure 1. The yields are rather different in the three sets of models. The decrease of $Z_{\pi^{ch}}$ at incident energies above 10 GeV in the calculations of HKHM [6] and BGS [5] is correlated with the onset of kaon production. The curves of the BN calculation [4] have a different behaviour, reflecting less production of low energy pions (see Fig. 5 below). This key difference is directly related to the hardness of the BN neutrino energy spectrum in Table 1.

In addition to the differences in energy-dependence and magnitude of the Z-factors, there are also some significant differences in the charge ratios. For almost all of the relevant range of energies, the ratio $Z_{p \to \pi^+}/Z_{p \to \pi^-}$ is significantly larger for BN than for either FRITIOF (HKHM $E_N > 5$ GeV) or TARGET (BGS), as shown in Fig. 2. The 5 GeV value of NUCRIN (HKHM) is intermediate, and at lower energy NUCRIN gives a very high value of the ratio. This is an artifact of NUCRIN, which neglects the charge exchange process $p + n \to \Delta^0 + p$, followed by $\Delta^0 \to p\pi^-$. These differences are directly relevant to the differences in the ratio $R_e = \bar{\nu}_e/\nu_e$ in the three calculations.

To estimate the effect of these differences on the neutrino flux, we plot the yield of charged pions in three different momentum windows, weighted by the primary spectrum $E_N^{-1.7}$. This is shown in Fig. 3 for pion momenta 0.3–0.4 GeV/$c$, 3–4 GeV/$c$ and 6–8 GeV/$c$. The area under the curves is proportional to the pion flux in the three momentum windows, which are chosen to reflect three different characteristic neutrino energies: $\sim$ 100 MeV, $\sim$ 1 GeV and $\sim$ 2 GeV respectively. The ratios BN/BGS for the three energy bands is respectively $\sim 0.3$, $\sim 0.6$, and $\sim 0.8 – 0.9$. These roughly correspond to the ratios of BN/BGS in Table 1. We therefore conclude that the difference in treatment of pion production in proton-air interactions is the main source of the difference between these two flux calculations.

To explore further the differences between the interaction models we compare in Fig. 4 the total multiplicity of charged pions in the three representations of hadronic interactions to data on proton-proton collisions [31]. Both BGS and HKHM have significantly higher multiplicity than BN, which is similar to $pp$ collisions. Both BGS and HKHM have a multiplicity on nuclear targets (A=14.5) that is about 50% higher than for proton targets. The pion production in the three interaction models is compared with data on light nuclei (beryllium) in Fig. 5. This data [28,29] is for beam momenta in the range 19-24 GeV/$c$, which is the median energy for production of $\sim$ GeV neutrinos [30]. All three models fit the data well for $x \gtrsim 0.2$. At smaller $x$ BN has much lower pion yield than the other two models, which leads to a correspondingly low result for the calculated neutrino flux below 1 GeV. The difference becomes progressively less with increasing neutrino energy because the representations of pion production agree with each other rather well at higher pion energy.

Much of the available accelerator data in the relevant range of beam momentum for protons incident on light nuclei is from experiments carried out to calculate accelerator neutrino beams. For example, the Eichten et al. experiment [29] was performed to provide data for calculation of the neutrino beam at the CERN-PS. A survey used at Brookhaven is that of Sanford & Wang [32]. Data used to calculate the Argonne neutrino beam is summarized by Barish et al. [33]. In all these cases the data are for relatively high pion momentum, so the ambiguity at low momenta in Fig. 5 is difficult to resolve. If we use data on proton targets at 24 GeV [34] to extend the data into the region of $x < 0.15$, then the data would favor the model used by BN.[3] We emphasize, however, that what is relevant is proton interactions in nitrogen and oxygen nuclei, in which the multiplicity of low energy pions is bound to be enhanced to some degree. The model of BGS and the Lund FRITIOF [22] model as used by HKHM both give significantly higher pion yields at low $x$ than that used by BN. We consider this question to be unresolved at present.

## IV. SUMMARY

We find that differences in the representation of the production of pions in $\sim 10$ to $\sim 30$ GeV interactions of protons with light nuclei are the major source of differences among three independent calculations of the flux of atmospheric neutrinos in the GeV energy range. Approximate treatment of the geomagnetic cutoffs also contributes significantly, especially for low energy neutrinos at Kamioka, which is the site with the highest geomagnetic cutoff for downward cosmic rays.

The lower neutrino flux below 1 GeV in the BN calculation is due to the representation of pion production they use, which is similar to interactions on a nucleon target. On the other hand, both BGS and HKHM use representations of proton-nucleus data that correspond to a pion multiplicity enhanced by a factor 1.6 as compared to proton-proton collisions.

The approximate representation of the geomagnetic cutoffs used by BGS tends to overestimate the intensity of low energy cosmic rays that penetrate through the geomagnetic field to the atmosphere. This is mainly because the effect of the Earth's penumbra has been neglected. The preferred approach is to use the ray-tracing technique, as done by HKHM and LK. In a new calculation of the geomagnetic cutoffs, including all details properly, Lipari and Stanev [21] find neutrino fluxes at 1 GeV reduced by factors of 0.85, 0.92 and 0.97 relative to BGS respectively at Kamioka, Gran Sasso/Frejus and IMB/Soudan. An updated version of BGS neutrino fluxes [12], based on the correct treatment of the geomagnetic cutoffs, predicts neutrino fluxes equal to, or slightly lower than, those of HKHM.

---

[3]We are grateful to D.H. Perkins for pointing out this reference to us and making this comparison [35].

We have referred earlier to the fact that the primary spectrum used by HKHM is significantly higher than that of the other two calculations. In addition, it shows a very large variation between solar maximum and solar minimum, even at equatorial latitudes. On the other hand, the primary spectrum of BGS [30] gives a solar cycle variation that is probably too small. The level of uncertainties associated with the primary spectrum is currently $\pm 10\%$. The normalization of the primary cosmic ray spectrum and its dependence on the solar cycle epoch should be a subject of future studies.

The differences in the $\bar{\nu}_e/\nu_e$ ratio, which are not yet a major factor in the interpretation of the experimental results, have to be further explored.

Comparison to high altitude muons may be used to check the normalization in a global way that may avoid the need to resolve all the various differences in the input to a calculation that starts from the primary cosmic ray spectrum [35]. There are recent measurements of muons at high altitude [36] to which calculations discussed in this paper can be compared (see e.g. Ref. [37]).

**Acknowledgements.** The research of TKG and TS is supported in part by the U.S. Department of Energy under DE-FG02-91ER40626.A007. The research of SM is supported in part by the Grant-in-Aid for Scientific Research, of the Ministry of Education, Science and Culture of Japan No. 07640419.

TABLES

TABLE I. Comparison of calculated neutrinos fluxes at Kamioka

|      | $\nu_\mu + \bar\nu_\mu$ | | | $\nu_e + \bar\nu_e$ | | | $\bar\nu_\mu/\nu_\mu$ | $\bar\nu_e/\nu_e$ | $R_{e/\mu}$ |
|------|------|------|------|------|------|------|------|------|------|
|      | $0.4-1$ | $1-2$ | $2-3$ | $0.4-1$ | $1-2$ | $2-3$ | $0.4$ | $\leq E_\nu$ | $\leq 1$ GeV |
| BGS  | 1.00 | 1.00 | 1.00 | 1.00 | 1.00 | 1.00 | 0.99 | 0.89 | 0.49 |
| HKHM | 0.90 | 0.95 | 1.04 | 0.87 | 0.91 | 0.97 | 0.99 | 0.84 | 0.48 |
| BN   | 0.63 | 0.79 | 0.95 | 0.62 | 0.74 | 0.87 | 0.98 | 0.76 | 0.50 |

TABLE II. Comparison of differential spectra (solar minimum) used by BGS and BN. (m$^{-2}$sr$^{-1}$s$^{-1}$GeV$^{-1}$; Energy is total energy per nucleon.)

| E (GeV) | free | | bound | |
|---|---|---|---|---|
|   | BGS | BN | BGS | BN |
| 2.0   | 802     | 957     | 536    | 241     |
| 3.17  | 318     | 332     | 164    | 83      |
| 5.02  | 126     | 114     | 51     | 28      |
| 7.96  | 42      | 39      | 15.4   | 9.4     |
| 12.6  | 12.8    | 13      | 4.6    | 2.93    |
| 20    | 3.8     | 4.3     | 1.33   | 0.89    |
| 31.7  | 1.12    | 1.36    | 0.40   | 0.25    |
| 50.2  | 0.33    | 0.45    | 0.115  | 0.071   |
| 79.6  | 0.098   | 0.129   | 0.034  | 0.019   |
| 126   | 0.029   | 0.037   | 0.0098 | 0.0048  |
| 200   | 0.0086  | 0.0104  | 0.0028 | 0.00122 |
| 317   | 0.0026  | 0.0029  | 0.00076| 0.00029 |
| 502   | 0.00076 | 0.00083 | 0.00023| 0.00007 |

TABLE III. Comparison of differential spectra (solar average) used by BGS and HKHM. ($m^{-2}sr^{-1}s^{-1}GeV^{-1}$; Energy is total energy per nucleon.)

| E | free | | bound | |
|---|---|---|---|---|
| (GeV) | BGS | HKHM | BGS | HKHM |
| 2.0 | 468 | 660 | 387 | 254 |
| 3.17 | 226 | 311 | 126 | 102 |
| 5.02 | 93.5 | 116 | 45 | 34 |
| 7.96 | 35 | 45.7 | 15.4 | 10.6 |
| 12.6 | 12.8 | 14.4 | 4.6 | 3.2 |
| 20 | 3.8 | 4.3 | 1.33 | 1.0 |
| 31.7 | 1.12 | 1.39 | 0.40 | 0.30 |
| 50.2 | 0.33 | 0.43 | 0.115 | 0.10 |
| 79.6 | 0.098 | 0.125 | 0.034 | 0.03 |
| 126 | 0.029 | 0.045 | 0.0098 | 0.009 |
| 200 | 0.0086 | 0.0127 | 0.0028 | 0.0026 |
| 317 | 0.0026 | 0.0044 | 0.00076 | 0.00064 |
| 502 | 0.00076 | 0.0014 | 0.00023 | 0.00017 |

FIGURES

FIG. 1. Z-factors for charged pions in proton-air collisions. Solid: BGS [5]; dotted: HKHM [6]; dashed: BN [4].

FIG. 2. Ratio of Z-factors for $\pi^+$ and $\pi^-$. (Same codes as Fig. 1).

FIG. 3. Average number of charged pions produced by a nucleon with total energy $E_0$ incident on air (multiplied by $E_0^{-1.7}$). Results are shown for three different bins of pion momentum for BGS [5] (solid lines) and for BN [4] (dashed lines).

FIG. 4. Average multiplicity of charged pions per interaction of proton ($E_p$ = total energy) in air (except for line which is for proton-proton collisions from Ref. [31]).

FIG. 5. Distributions of fractional momentum $(\mathrm{d}n\,/\,\mathrm{d}\ln x)$ of charged pions produced in interactions of $\approx 20$ GeV/$c$ momentum protons with light nuclei. Models are shown for target = air, data [28,29] for target = Be. $\pi^+$ and $\pi^-$ are shown separately.

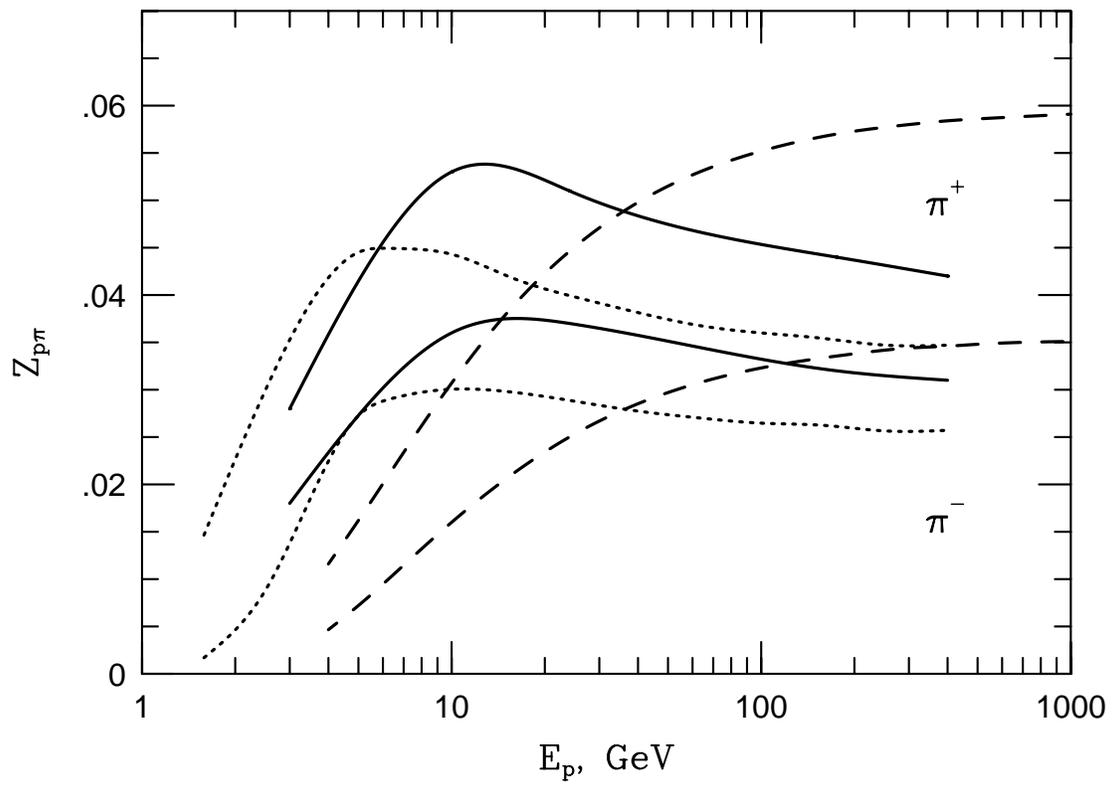

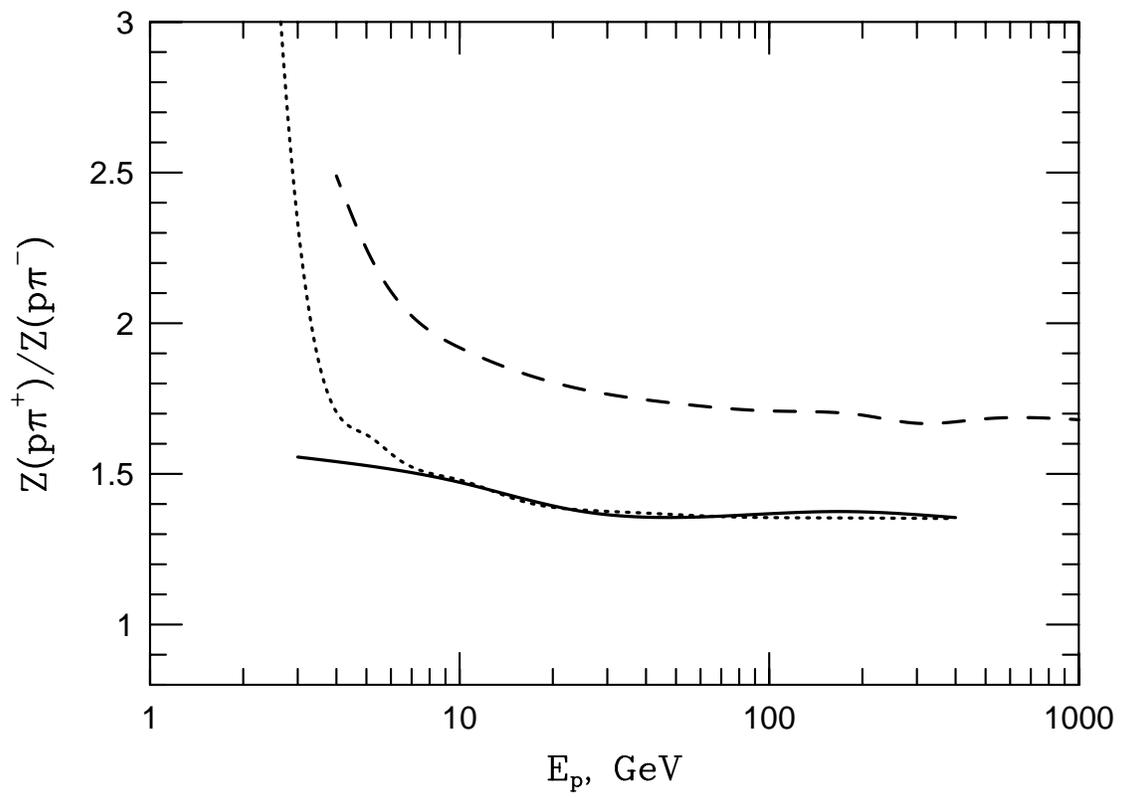

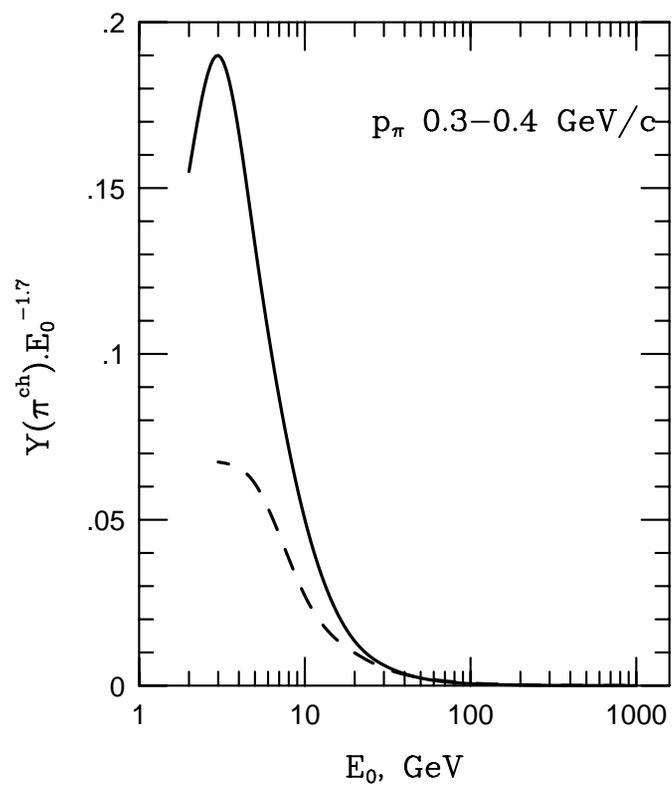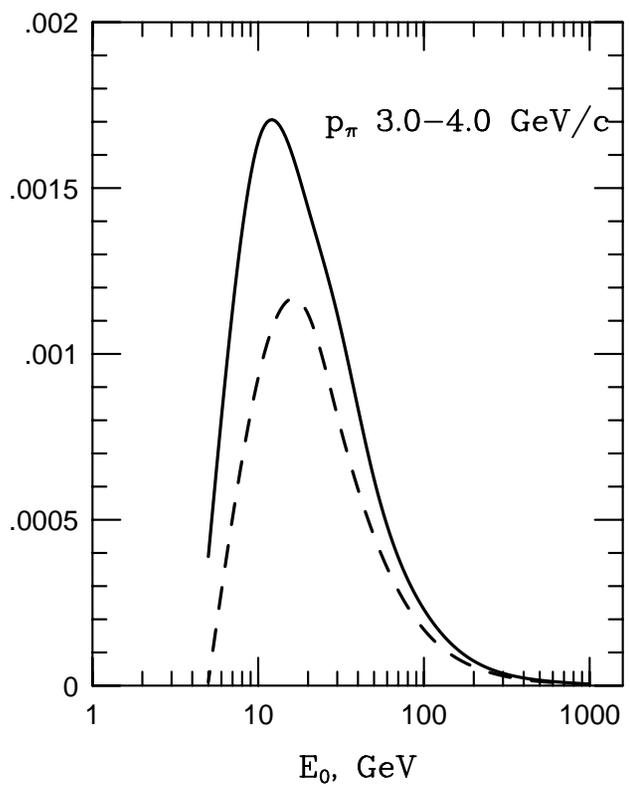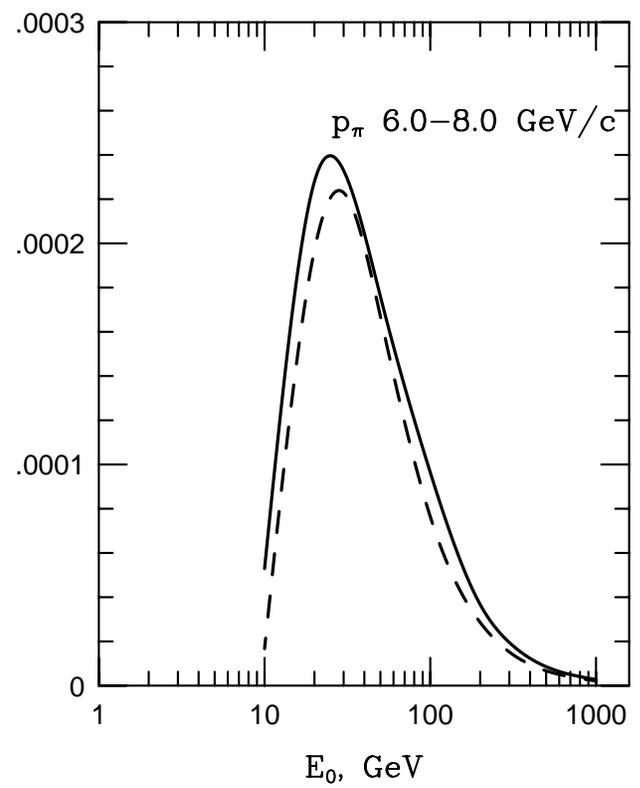

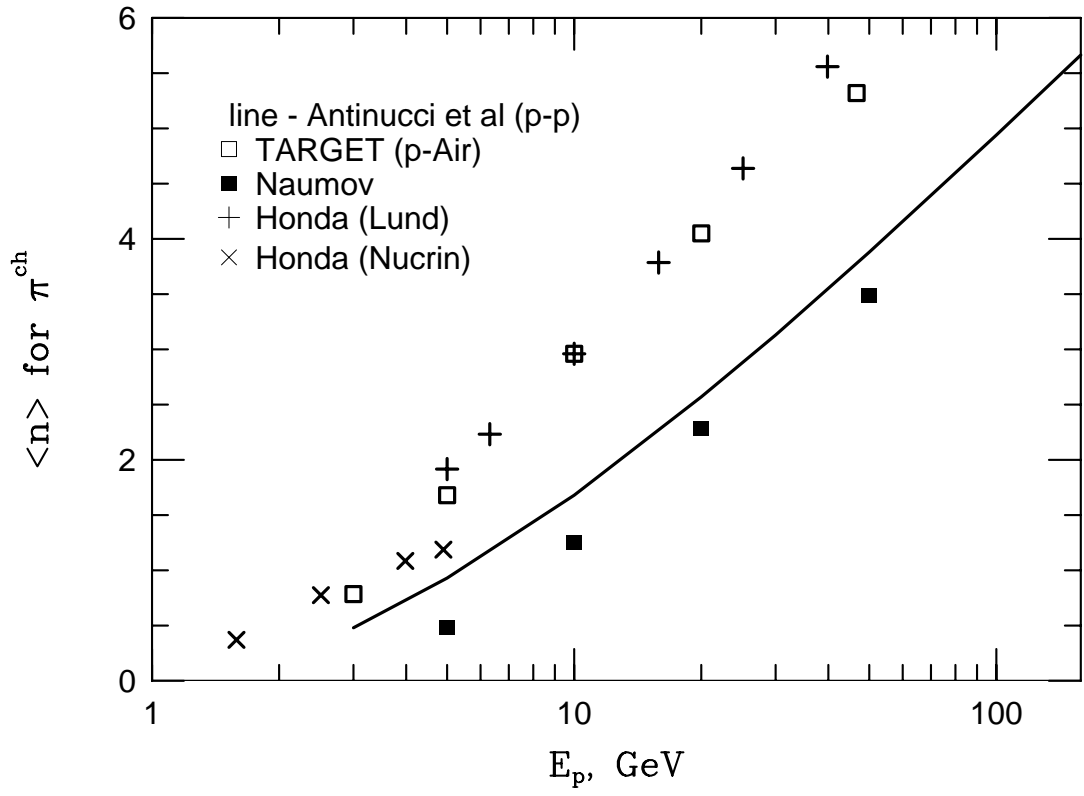

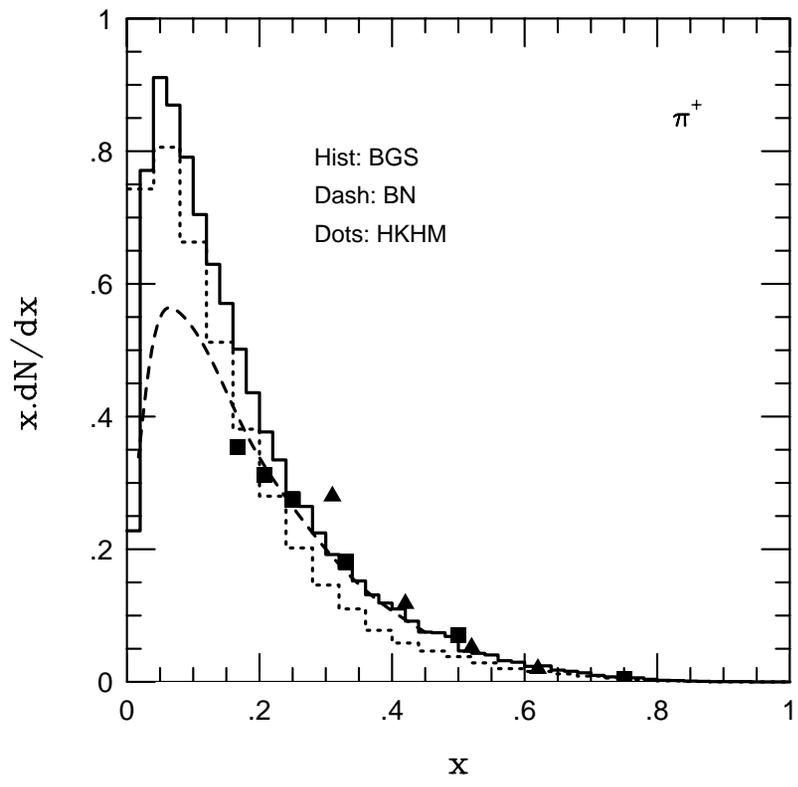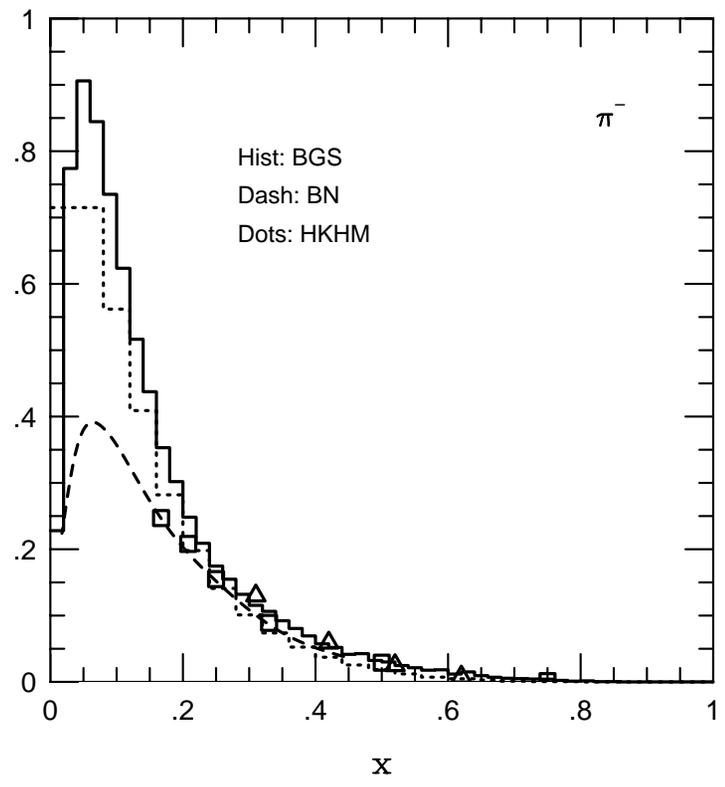